\pdfoutput=1
\documentclass[pra,aps,10pt,a4paper,nofootinbib,notitlepage,onecolumn,superscriptaddress]{revtex4-1}
\usepackage[utf8]{inputenc}
\usepackage[T1]{fontenc}
\usepackage[sc,osf]{mathpazo}
\usepackage[scaled=0.86]{berasans}
\usepackage[scaled=1.03]{inconsolata}
\usepackage[british]{babel}
\usepackage[unicode,colorlinks]{hyperref}
\usepackage{amsmath,amssymb,amsthm}
\usepackage{xspace}
\usepackage{mathtools}

\DeclareMathOperator{\tr}{tr}

\let\originalleft\left
\let\originalright\right
\renewcommand{\left}{\mathopen{}\mathclose\bgroup\originalleft}
\renewcommand{\right}{\aftergroup\egroup\originalright}

\newcommand{\KetBra}[1]{{\Ket{#1}\!\Bra{#1} }}

\newcommand{\Bra}[1]{{ \langle \! \langle{#1}\vert }}
\newcommand{\Ket}[1]{{ \vert {#1}  \rangle \!  \rangle}}

\newcommand{\bra}[1]{\langle #1|}

\newcommand{\ket}[1]{| #1 \rangle}

\newcommand{\ketbra}[1]{\left|#1\middle\rangle\middle\langle#1\right|}

\newcommand{\norm}[1]{\left\|#1\right\|}

\newcommand{\abs}[1]{\left|#1\right|}

\newcommand{\de}[1]{\left(#1\right)}
\newcommand{\deBig}[1]{\Big(#1\Big)}
\newcommand{\De}[1]{\left[#1\right]}
\newcommand{\DeBig}[1]{\Big[#1\Big]}

\newcommand{\bqplctc}{\ensuremath{\text{\textup{BQP}}_{\ell\text{\textup{CTC}}}}\xspace}
\newcommand{\mathand}{\quad\text{and}\quad}

\newcommand{\id}{\mathbb{1}}

\newcommand{\be}{\begin{equation}}
\newcommand{\ee}{\end{equation}}

\newcommand{\ie}{i.e.\@\xspace}

\newtheorem{theorem}{Theorem}
\newtheorem*{theorem*}{Theorem}

\newtheorem{definition}[theorem]{Definition}

\mathtoolsset{centercolon}

\input{Qcircuit}
   
\hypersetup{pdftitle={{Quantum computation with indefinite causal structures}}}

\begin{document}
\title{Quantum computation with indefinite causal structures}
\author{Mateus Araújo}
\affiliation{Institute for Theoretical Physics, University of Cologne, Zülpicher Straße 77, 50937 Cologne, Germany}
\author{Philippe Allard Guérin}
\affiliation{Faculty of Physics, University of Vienna, Boltzmanngasse 5, 1090 Vienna, Austria}
\affiliation{Institute for Quantum Optics and Quantum Information (IQOQI), Austrian Academy of Sciences, Boltzmanngasse 3, 1090 Vienna, Austria}
\author{Ämin Baumeler}
\affiliation{Faculty of Informatics, Università della Svizzera italiana, Via G. Buffi 13, 6900 Lugano, Switzerland}
\affiliation{Facoltà indipendente di Gandria, Lunga scala, 6978 Gandria, Switzerland}
\date{\today}

\begin{abstract}
One way to study the physical plausibility of closed timelike curves (CTCs) is to examine their computational power. This has been done for Deutschian CTCs (D-CTCs) and post-selection CTCs (P-CTCs), with the result that they allow for the efficient solution of problems in PSPACE and PP, respectively. Since these are extremely powerful complexity classes, which are not expected to be solvable in reality, this can be taken as evidence that these models for CTCs are pathological. This problem is closely related to the nonlinearity of this models, which also allows for example cloning quantum states, in the case of D-CTCs, or distinguishing non-orthogonal quantum states, in the case of P-CTCs. In contrast, the process matrix formalism allows one to model indefinite causal structures in a \emph{linear} way, getting rid of these effects, and raising the possibility that its computational power is rather tame. In this paper we show that process matrices correspond to a linear particular case of P-CTCs, and therefore that its computational power is upperbounded by that of PP. We show, furthermore, a family of processes that can violate causal inequalities but nevertheless can be simulated by a causally ordered quantum circuit with only a constant overhead, showing that indefinite causality is not necessarily hard to simulate.
\end{abstract}

\maketitle

\section{Introduction}

Gödel found in 1949 a spacetime allowed by general relativity that features closed timelike curves (CTCs) \cite{godel49} (see also \cite{lanczos24}), awakening bafflement and interest from the physical community. CTCs have since then been intensely studied from the general relativistic point of view \cite{morris88,novikov89,hawking92}, without a clear conclusion about whether they are actually possible. To separate the technical difficulties of general relativity from intrinsic properties of CTCs, Deutsch proposed in 1991 to abstract away the spacetime geometry, and study them from the point of view of the theory of quantum computation and information instead \cite{deutsch91}. 

His approach proved to be rather fruitful, allowing for the discovery of several new counterintuitive properties of CTCs. Specifically, Deutsch's CTCs (D-CTCs) were shown to be non-unitary, non-linear, to be able to distinguish non-orthogonal states \cite{brun09}, to clone quantum states \cite{ahn13,brun13}, to violate Heisenberg's uncertainty principle \cite{pienaar13}, and to allow for the efficient solution of problems in PSPACE \cite{aaronson09}. Of course, it is possible that these features are not counterintuitive properties of real CTCs (if they exist), but rather pathologies of Deutsch's model. Indeed, a somewhat less problematic model was later introduced by several authors \cite{politzer94,bennett05,svetlichny11}, based on post-selected teleportation (more details below). Their post-selection CTCs (P-CTCs) are still non-linear, though, able to distinguish linearly independent non-orthogonal states \cite{brun12}, and able to efficiently solve problems in PP \cite{lloyd11,brun12}. Even though PP is believed to be strictly contained in PSPACE, it is still an extremely powerful complexity class, including problems even harder than NP-complete ones. If the efficient solution of NP-complete problems turns out not to be possible, as it is widely suspected \cite{aaronson05b}, even the milder contradictions with standard quantum mechanics generated by P-CTCs would be just pathologies of the model.

In contrast with D-CTCs and P-CTCs, the process matrix formalism is a \emph{linear} extension of quantum mechanics that is able to model situations in which there is no definite causal order between events \cite{oreshkov12}. It includes even a non-trivial unitary particular case \cite{araujo16}. One can therefore expect its departure from quantum mechanics to be correspondingly less dramatic. While some process matrices are strongly acausal, as shown by the violation of causal inequalities \cite{oreshkov12,baumeler13,baumeler14,branciard15simplest,baumeler16,oreshkov15,abbott16}, the partial results known about their computational power indicate that it is not implausibly large: the biggest advantage known over quantum computers is a reduction in query complexity from $O(n^2)$ to $O(n)$, enabled by the quantum switch \cite{chiribella09,chiribella12,araujo14}. Furthermore, the power of classical deterministic processes was shown to be upperbounded by UP $\cap$ coUP \cite{baumeler16c,baumeler16d,baumeler16e}. This latter class is contained in NP and consists of all decision problems where, for a given instance, membership or non-membership can be proven efficiently via a \emph{unique} witness. This means that classical deterministic processes cannot solve NP-complete problems unless \mbox{UP $\cap$ coUP $=$ NP}, which is highly doubted. While potentially disappointing, this lack of drama makes process matrices more likely to be physical than D-CTCs or P-CTCs. This point is underscored by the fact that the quantum switch has already been implemented experimentally, without using any exotic physics \cite{procopio_experimental_2014,rubino16} (see also Ref.~\cite{baumeler17} for a comparison between process matrices, D-CTCs, and P-CTCs). 

Here we show that process matrices correspond to a linear particular case of P-CTCs, and use this equivalence to motivate the definition of a complexity class for the problems that can be efficiently solved by process matrices: \bqplctc. It immediately follows that $\bqplctc \subseteq \text{PP}$. We also exhibit a family of process matrices that, unlike the quantum switch \cite{araujo15witnessing}, can always violate causal inequalities but can nevertheless be simulated by a quantum circuit with merely a constant overhead. This shows that less causality does not imply more computational power.

\section{Post-selection closed timelike curves}

The model of closed timelike curves we use was proposed independently by several authors \cite{politzer94,bennett05,svetlichny11} (see also \cite{dias11,lloyd11,brun12}). It is motivated by the observation that, in the quantum teleportation protocol, one does not need to do any correction on the target state when the outcome of the Bell measurement is $\ket{\phi^+}$.
In this case one can write the target state as already being there before the measurement was made, and even before the state to be teleported was created, as shown in the following circuit:
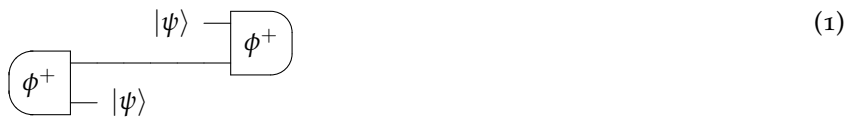
\begin{equation}\label{eq:teleportation}
 \Qcircuit @C=1em @R=.6em @!R  {
 & & 	&  & & \lstick{\ket{\psi}} &   \multimeasureD{1}{\phi^+}\\
\multiprepareC{1}{\phi^+} &  \qw  & \qw  &\qw &\qw & \qw & \ghost{\phi^+}   \\
\pureghost{\phi^+}    & \rstick{\ket{\psi}} \qw  &&  &  &   & & \\
}
\end{equation}
This post-selected teleportation protocol is then used to teleport the output state of a part of the output of a unitary matrix to a part of its input, whereas the rest of the unitary simply maps a quantum system from the past $P$ to the future $F$, as shown in the circuit \eqref{eq:post-selection}.
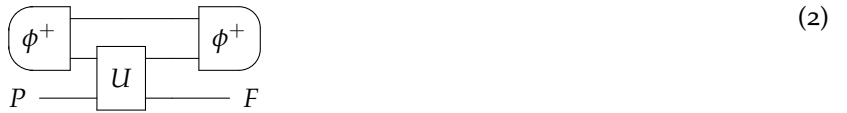
\begin{equation}\label{eq:post-selection}
 \Qcircuit @C=1em @R=.6em @!R  {
\multiprepareC{1}{\phi^+} &  \qw  & \qw &   \multimeasureD{1}{\phi^+} \\
\pureghost{\phi^+}    &\multigate{1}{U}  &\qw  & \ghost{\phi^+}  \\
  \lstick{P}  & \ghost{U} & \qw &\rstick{F} \qw     &
}
\end{equation}
Naming the subsystems as $1,2$, and $3$ (from bottom to top), the operator that maps the chronology-respecting system from the past to the future is then
\be
K = \bra{\phi^+}^{23} U^{12} \ket{\phi^+}^{23} \propto \tr_2 U^{12}.
\ee
Since a partial trace of a unitary is in general not a unitary, $K$ will not necessarily preserve the norm of a input state, so for the evolution to be valid we will need to renormalize it. The evolution of a state $\ket{\psi}$ input in $P$ is therefore defined as
\be\label{eq:ctc_normalization}
\ket{\psi} \mapsto \frac{1}{\norm{K \ket{\psi}}}_2K \ket{\psi},
\ee
which is undefined when $\norm{K \ket\psi}_2 = 0$, illustrating how pathological this model is.

This post-selected closed timelike curve (known as P-CTC) is usually represented as
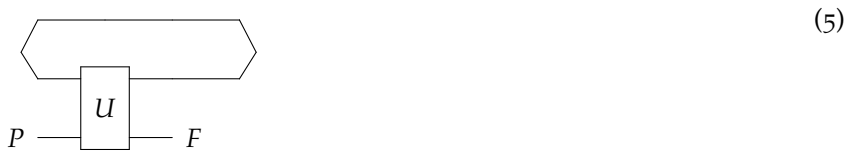
\begin{equation}
 \Qcircuit @C=.7em @R=.4em @!  {
\preselection{1}{1.0} &  \qw  & \qw &  \qw \postselection{1}{1.0} \\
    &\multigate{1}{U}  &\qw  & \qw   & \\
  \lstick{P}  & \ghost{U} & \rstick{F} \qw     &
}
\end{equation}
with the renormalization implicitly included.

Note that although we used pure states and unitary transformations, the formalism of P-CTCs can be easily extended to mixed states and CPTP maps.

\section{Processes as linear closed timelike curves}\label{sec:equivalence}

Now we proceed to show how having access to P-CTCs allows one to implement any process, or equivalently how to simulate any process using post-selected teleportation. It was already known that the bipartite quantum switch can be implemented using one P-CTC \cite{chiribella09}, and that one can simulate any process matrix via post-selection \cite{oreshkov16,costa14private,silva17}. What is novel here is the generality of the connection with P-CTCs, which gives us both a nice interpretation for the post-selection and a convenient representation for studying computational complexity.

We shall use the definition of processes from Ref.~\cite{araujo16}, which says that an operator $W \in P \otimes F \otimes \bigotimes_{k=0}^{n-1} A^k_I \otimes A^k_O$ is a process matrix if and only if for all CPTP maps $\mathcal A^k : A^k_I\otimes {A^k_I}' \to A^k_O \otimes {A^k_O}'$ the operator
\begin{equation}\label{eq:definition_w}
 G = \tr_{\bigotimes_{k=0}^{n-1}A^k_IA^k_O} \De{W^{T_{\bigotimes_{k=0}^{n-1}A^k_IA^k_O}}\de{ \bigotimes_{k=0}^{n-1} {A^k}} }
\end{equation}
is a valid CPTP map from $P\otimes \bigotimes_{k=0}^{n-1}{A^k_I}'$ to $F\otimes \bigotimes_{k=0}^{n-1}{A^k_O}'$, where identity matrices are left implicit in this product, $\cdot^{T_{\bigotimes_{k=0}^{n-1}A^k_IA^k_O}}$ denotes partial transposition over $\bigotimes_{k=0}^{n-1}A^k_IA^k_O$, and
\begin{gather}
 A^k = \mathcal I \otimes {\mathcal A^k} (\KetBra{\id^{A^k_I{A^k_I}'}}) \in A^k_I \otimes {A^k_I}' \otimes A^k_O \otimes {A^k_O}' \\
 G = \mathcal I \otimes {\mathcal G} (\KetBra{\id^{P\bigotimes_{k=0}^{n-1}{A^k_I}'}}) \in P \otimes \bigotimes_{k=0}^{n-1} {A^k_I}' \otimes F \otimes \bigotimes_{k=0}^{n-1} {A^k_O}'
\end{gather}
are the Choi-Jamiołkowski (CJ) representations of the CPTP maps $\mathcal A^k$ and $\mathcal G$, with the ``double-ket'' of a $a \times b$ matrix $M$ being defined as
$\Ket{M} = \sum_{i=0}^{b-1} \ket{i}M\ket{i}$. A more thorough review of the CJ isomorphism can be found in Refs.~\cite{chiribella09b,araujo16}.

To reduce clutter, we shall present only the bipartite version of the argument, renaming $A^0$ as $A$ and $A^1$ as $B$. We shall also take the dimension of the auxiliary Hilbert spaces $A_I'$, $A_O'$, $B_I'$, and $B_O'$ to be one. It is straightforward to do away with such restrictions.

With these simplifications, equation \eqref{eq:definition_w} reduces to
\begin{equation}
 G_{A,B} = \tr_{A_IA_OB_IB_O} \De{W  (A^T\otimes B^T) }.
\end{equation}
Substituting the definition of the CJ isomorphism for $A$ and $B$ and noting that
\be
\de{\mathcal I \otimes {\mathcal A} (\KetBra{\id^{A_I}})}^T = \mathcal A^* \otimes I (\KetBra{\id^{A_O}}),
\ee
where $\mathcal A^*$ is the adjoint of $\mathcal A$, we arrive at
\be
G_{A,B} = \tr_{A_IA_OB_IB_O} \De{W \de{{\mathcal A}^* \otimes \mathcal I (\KetBra{\id^{A_O}})} \otimes \de{{\mathcal B}^* \otimes \mathcal I (\KetBra{\id^{B_O}})} },
\ee
which after collecting terms becomes
\be\label{eq:gab}
G_{A,B} = \tr_{A_IA_OB_IB_O} \De{W \deBig{({\mathcal A}^*\otimes {\mathcal B}^*) \otimes \mathcal I (\KetBra{\id^{A_OB_O}})} }.
\ee
As noted in Ref.~\cite{araujo16}, $W$ must be itself the CJ representation of a CPTP map $\mathcal W$ from $P \otimes A_O \otimes B_O$ to $F \otimes A_I \otimes B_I$, so defining it via 
\be
W = \mathcal W \otimes \mathcal I (\KetBra{\id^{A_OB_OP}})
\ee
and substituting this in equation \eqref{eq:gab} shows us that $G_{A,B}$ is equal to
\be
\tr_{A_IA_OB_IB_O}\DeBig{ \deBig{{\mathcal W \otimes \mathcal I (\KetBra{\id^{A_OB_OP}})}} \deBig{({\mathcal A}^*\otimes {\mathcal B}^*) \otimes \mathcal I (\KetBra{\id^{A_OB_O}})}}
\ee
which, after taking the adjoint of the map ${\mathcal A}^*\otimes {\mathcal B}^*$ to the other side of the Hilbert-Schmidt inner product, becomes
\be
\tr_{A_IA_OB_IB_O}\DeBig{ \deBig{\big((\mathcal A \otimes \mathcal B) \circ \mathcal W\big) \otimes \mathcal I (\KetBra{\id^{A_OB_OP}})} \KetBra{\id^{A_OB_O}}}
\ee
Finally, we replace $\Ket{\id^{A_OB_O}}$ with the non-normalized maximally entangled state $\sqrt{d_{A_O}d_{B_O}}\ket{\phi^+}^{A_OB_O}$, which gives us
\be
G_{A,B} = d_{A_O}^2d_{B_O}^2 \tr_{A_IA_OB_IB_O}\bigg[\deBig{\big((\mathcal A \otimes \mathcal B) \circ \mathcal W\big) \otimes \mathcal I \de{\KetBra{\id^P}\otimes\ketbra{\phi^+}^{A_OB_O}}} \ketbra{\phi^+}^{A_OB_O} \bigg]
\ee
This identity can be stated more clearly as the circuit equation:
\begin{equation}\label{eq:wpost-selection_circuit}
\mathcal G_{\mathcal A,\mathcal B} = d_{A_O}^2d_{B_O}^2\ \vcenter{\hbox{\Qcircuit @C=0.3em @R=0.3em @!R {
 \multiprepareC{3}{\phi^+} & \qw                       & \qw                       & \qw                 & \qw                       & \multimeasureD{3}{\phi^+} \\
                           & \multiprepareC{1}{\phi^+} & \qw                       & \qw                 & \multimeasureD{1}{\phi^+} &                           \\
                           & \pureghost{\phi^+}        & \multigate{2}{\mathcal W} & \gate{\mathcal A}   & \ghost{\phi^+}            &                           \\
 \pureghost{\phi^+}        & \qw                       & \ghost{\mathcal W}        & \gate{\mathcal B}   & \qw                       & \ghost{\phi^+}            \\
  	        	   & \lstick{P}		       & \ghost{\mathcal W}        & \rstick{F} \qw      &                           &                           \\
}}}
\end{equation}
which shows how to simulate the map $\mathcal G_{\mathcal A,\mathcal B}$ via post-selection: one prepares the state $\ket{\phi^+}^{A_OB_O} = \ket{\phi^+}^{A_O} \otimes \ket{\phi^+}^{B_O}$, applies the CPTP map $\mathcal W$ to it, applies the CPTP maps $\mathcal A$ and $\mathcal B$ to the output of $\mathcal W$, and measures the resulting state in a basis that includes $\ket{\phi^+}^{A_OB_O}$, post-selecting on obtaining it.

In the language of P-CTCs, one can also understood this as applying the CPTP maps $\mathcal A$ and $\mathcal B$ to the output of $\mathcal W$, and using a P-CTC to teleport the output of $\mathcal A$ and $\mathcal B$ to the input of $\mathcal W$. It can be represented more compactly as
\begin{equation}\label{eq:w_ctc_circuit}
\mathcal G_{\mathcal A,\mathcal B} = \vcenter{\hbox{\Qcircuit @C=1em @R=0.3em @!R {
 \preselection{3}{2.4} & \qw                       & \qw               & \qw \postselection{3}{2.4} \\
 \preselection{1}{0.8} & \qw                       & \qw               & \qw \postselection{1}{0.8} \\
                               & \multigate{2}{\mathcal W} & \gate{\mathcal A} & \qw                                \\
			       & \ghost{\mathcal W}        & \gate{\mathcal B} & \qw                                \\
		\lstick{P}     & \ghost{\mathcal W}        & \rstick{F} \qw    &                                    \\
}}}
\end{equation}
 
 The crucial point to notice is that the probability of post-selection is always $1/d_{A_O}^2d_{B_O}^2$, independently of the CPTP maps $\mathcal A$ and $\mathcal B$ or the state $\rho$ input in $P$, which makes the renormalization of this circuit, and therefore the mapping $\mathcal A,\mathcal B,\rho \mapsto \mathcal G_{\mathcal A,\mathcal B}(\rho)$ a \emph{linear} transformation. One could, instead, let $\mathcal W$ be an arbitrary CPTP map, as in a general P-CTC, and define the transformation from $P$ to $F$ as done in equation \eqref{eq:ctc_normalization}:
 \be
 \rho \mapsto \frac{\mathcal G_{\mathcal A,\mathcal B}(\rho)}{\tr \mathcal G_{\mathcal A,\mathcal B}(\rho)},
 \ee
 This transformation is linear only if the denominator is a constant, that does not depend on $\mathcal A$, $\mathcal B$, and $\rho$. But this is the case only if $\mathcal G_{\mathcal A,\mathcal B}$ is proportional to a CPTP map, for all $\mathcal A$ and $\mathcal B$, which is simply the definition of a process, up to an irrelevant multiplicative constant.
 
 This shows that process matrices are exactly the linear particular case of P-CTCs, when those are defined with the variable gates $\mathcal A$ and $\mathcal B$. Since their nonlinearity is the source of major pathologies, such as the ability to solve NP-complete problems and to distinguish non-orthogonal states, one can expect the process matrices to be a less problematic case of P-CTCs, that however still allow for indefinite causal order.

 \section{The computational model}
 
 Now we are in a position to define the complexity class of the problems that can be efficiently solved with process matrices. We shall give two equivalent definitions, one based directly on the definition of a process matrix, and the other based on its interpretation in terms of P-CTCs. In both of them we shall use as basic element the operator 
 $W \in P \otimes F \otimes \bigotimes_{k=0}^{n-1} A^k_I \otimes A^k_O $ which, as above, shall be understood as the CJ representation of a CPTP map from $P\otimes \bigotimes_{k=0}^{n-1}{A^k_O}$ to $F\otimes \bigotimes_{k=0}^{n-1}A^k_I$, where $P,F$ correspond to the global past and future of the circuit, and the other $n$ subsystems are connected to arbitrary CPTP maps $A^k \in A^k_I \otimes A^k_O$ in a causally indefinite way (that is, by fiat as in equation \eqref{eq:definition_w} or by using a P-CTC as in equation \eqref{eq:w_ctc_circuit}), where we have set the dimension of the ancillary systems ${A^k_I}'$ and ${A^k_O}'$ to be one for simplicity. The result of this connection is then a circuit $G$, which accepts or rejects an input $x$:
  \be\label{eq:complexity_definition}
  G =  \tr_{ \bigotimes_{k=0}^{n-1}A^k_IA^k_O} \De{W\de{ \bigotimes_{k=0}^{n-1} {A^k}^T} } = \ \ \vcenter{\hbox{\Qcircuit @C=1em @R=0.3em @!R {
\preselection{4}{3} & \qw  &\qw & \qw & \qw & \qw \postselection{4}{3} \\
\preselection{1}{0.8} & \qw &\qw & \qw  & \qw & \qw\postselection{1}{0.8} \\
&  \multigate{3}{\mathcal W} & \qw & \gate{\mathcal A^0} & \qw& \qw \\
&  \pureghost{\mathcal W} & & \raisebox{0.7em}{\vdots} & \\
&  \ghost{\mathcal W} & \qw & \gate{\mathcal A^{n-1}} &\qw & \qw  \\
\lstick{P}  &  \ghost{\mathcal W} & \rstick{F} \qw
}}}
 \ee
 As long as one doesn't cheat by encoding hard-to-compute functions in the CPTP maps $A^k$ they can be chosen as convenience dictates. If concreteness is desired, one can fix them without loss of generality to be the identity map $A^k = \KetBra{\id^{A^k_I}}$, as this allows any other choice of CPTP maps to be absorbed into the operator $W$ itself. If $W$ is the CJ representation of a unitary, i.e., if $W = \KetBra{U_W}$ for some unitary $U_W$, this choice leads to a particularly simple formula for the unitary $U_G$ such that $G = \KetBra{U_G}$:
 \be
 U_G = \tr_{A^0_I\ldots A^{n-1}_I} U_W.
 \ee
 
 The output of the circuit is then determined by a measurement on one of the qudits in $F$. If the result of the measurement is $1$ we say that the circuit accepts input $x$, and if the result is $0$ we say that input $x$ was rejected. As we shall see shortly, the number of CTCs $n$ must depend on the size of the input $x$ for the complexity class to be nontrivial.
 
 The complexity class is then: 
\begin{definition}
 \bqplctc is the class of languages $L$ for which there exists a polynomial-time classical algorithm that constructs a polynomial-size quantum circuit $W$ such that for all inputs $x$ of size $m$:
 \begin{enumerate}
  \item For all CPTP maps $A^k$ the operator $G$ defined in equation \eqref{eq:complexity_definition} is a CPTP map.
  \item If $x \in L$ then $P[G\text{ accepts }x] \ge \frac23$.
  \item If $x \notin L$ then $P[G\text{ rejects }x] \ge \frac23$. 
 \end{enumerate}
\end{definition}
It should be clear, given the interpretation of processes in terms of P-CTCs shown in the previous section, that this definition is equivalent to:
\begin{definition}
 \bqplctc is the class of languages $L$ for which there exists a polynomial-time classical algorithm that constructs a polynomial-size quantum circuit $W$ such that for all inputs $x$ of size $m$:
 \begin{enumerate}
  \item For all CPTP maps $A^k$ the probability that all post-selections in equation \eqref{eq:complexity_definition} succeed is $1/d^{2n}$.
  \item If $x \in L$ then given that all post-selections succeed $P[G\text{ accepts }x] \ge \frac23$.
  \item If $x \notin L$ then given that all post-selections succeed $P[G\text{ rejects }x] \ge \frac23$. 
 \end{enumerate}
\end{definition}
Note that although this computational model fits naturally with the notion of pure processes defined in \cite{araujo16}, we are not demanding $W$ to be the CJ representation of a unitary (corresponding to a pure process). It would be interesting to know whether demanding this would reduce the computational power of the model. It is vital, however, to allow the maps $A^k$ to be general CPTP maps, because if we restricted them to be unitaries we would get a larger set of circuits $W$ such that $G$ is a valid CPTP map (see Appendix C of Ref.~\cite{araujo16}). These extra circuits would be precisely the ones for which $G$ is not a valid CPTP map when the $A^k$ are non-unital maps. Since the process matrix formalism is based on the assumption that the maps $A^k$ represent local laboratories where normal quantum mechanics apply these extra circuits would be rather contrived, as natural models of noise can turn unitary maps into non-unital ones.

It should be immediately clear that $\text{BQP} \subseteq \bqplctc$, as one obtains a normal BQP circuit simply by setting $n = 0$ or by simulating a causally ordered circuit using the P-CTCs. 

Another consequence of this definition is that for \bqplctc to have any advantage over BQP, the number of P-CTCs $n$ must increase faster than logarithmically with the size of the input $m$. This is because one can simply repeat the circuit until all post-selections succeed, and if the number of repetitions is small enough then the repeated circuit will still be in BQP. More precisely, the probability that all post-selections succeed simultaneously is $d^{-2n}$, so repeating the circuit a number of times greater than $\log(10)d^{2n}$ guarantees that the probability that at all post-selections succeed simultaneously at least once is greater than $9/10$. The probability that all post-selections succeed at least once and that in this case the circuit accepts is then greater than $6/10$. Since the size of the repeated circuit is polynomial in $m$ if $n$ grows logarithmically with $m$ or slower, it is simply a valid BQP circuit. 

This stands in stark contrast with the case of general (non-linear) P-CTCs, where having access to a single one is already enough to access the power of the whole complexity class \cite{lloyd11,brun12}. The argument is simply that the power of quantum computation with P-CTCs is the same as the power of quantum computation with post-selection, and that a single P-CTC is sufficient to simulate any desired post-selection. Since post-selection is also sufficient to simulate the whole \bqplctc, it is clear that the complexity class of quantum computation with unrestricted post-selection -- called PostBQP -- contains \bqplctc. In its turn, PostBQP was shown by Aaronson \cite{aaronson05} to be equal to the rather powerful complexity class PP \cite{arora09}, which can be seen as a pathological version of BPP, that does not require the probabilities of acceptance and rejection to be bounded away from $1/2$.

Summarizing the conclusions of the previous paragraph, we have that\footnote{A technical detail is that for the equality $\text{PostBQP} = \text{PP}$ to hold one needs to restrict the post-selection probabilities to be exponentially small or greater (when nonzero) \cite{aaronson14}. Since this restriction is automatically obeyed by \bqplctc we don't need to worry about it.}
\[ \bqplctc \subseteq \text{PostBQP} = \text{PP}, \]
which is a rather unsatisfactory upper bound, as PP allows for example the solution of NP-complete problems, which we don't expect to be possible in a physically realizable computational model. A somewhat tighter upper bound is the class $\text{PostBQP}_\text{exp}$ defined by Morimae and Nishimura \cite{morimae15}, which is PostBQP with the restriction that the probability of successful post-selection must be equal to $2^{-q(m)}$ for some polynomial $q$. Since this is the case for \bqplctc, we have that
\[ \bqplctc \subseteq \text{PostBQP}_\text{exp} \subseteq \text{PostBQP} .\]
It is, however, unlikely that $\bqplctc = \text{PostBQP}_\text{exp}$, as the extra restriction that for \bqplctc the probabilities of post-selection must be equal to $d^{-2n}$ for all CPTP maps $A^k$ is in general not obeyed by $\text{PostBQP}_\text{exp}$. As an example, take $U_W = \id \otimes U^{\otimes n}$ and $A^k = \KetBra{V}$ for all $k$. The probabilities of post-selection are then $\frac{1}{d^{2n}}\abs{\tr UV}^{2n}$, which means that if $V=\id$ and $\abs{\tr U} = 1$ this is a valid $\text{PostBQP}_\text{exp}$ circuit. On the other hand, this is not a valid \bqplctc circuit, as one can make these probabilities be anything in $[0,1]$ by varying $V$.

\section{The quantum switch}

It might be enlightening to see concrete examples of causally nonseparable processes written in the P-CTC representation. A well-known one is the quantum switch \cite{chiribella09,chiribella12,araujo14}, which for $n$ parties is written as the process vector \cite{araujo16} 
\be\label{eq:wswitchket}
\ket{w_{\text{switch}n}} = \sum_{x=0}^{n!-1}\ket{x}^{P_1}\Ket{\id}^{P_2A^{\sigma_x(0)}_I}\de{\bigotimes_{k=0}^{n-2}\Ket{\id}^{A^{\sigma_x(k)}_O A^{\sigma_x(k+1)}_I}}\Ket{\id}^{A^{\sigma_x(n-1)}_OF_2}\ket{x}^{F_1},
\ee
where $\sigma_x$ is any labelling of permutations (for concreteness, it can be taken to be the one described in Appendix \ref{sec:nswitch}). A process vector is simply a convenient way to write rank-one (and in particular pure) process matrices, which can then be recovered by taking the outer product of the process vector with itself. For example, the process matrix of the quantum switch is given by
\be
W_{\text{switch}n} = \ketbra{w_{\text{switch}n}}.
\ee
For $n=2$ this process represents a quantum channel from Alice to Bob if a control system in the state $\ket{0}$ is input into $P_1$, and a quantum channel from Bob to Alice if the control system is in the state $\ket{1}$. The vector $\ket{w_{\text{switch}2}}$ is the CJ representation of a unitary from the spaces $P_1,P_2,A_O,B_O$ to the spaces $F_1,F_2,A_I,B_I$, so writing this unitary as a circuit allows us to represent this process in the P-CTC representation defined in equation \eqref{eq:w_ctc_circuit}:
\be
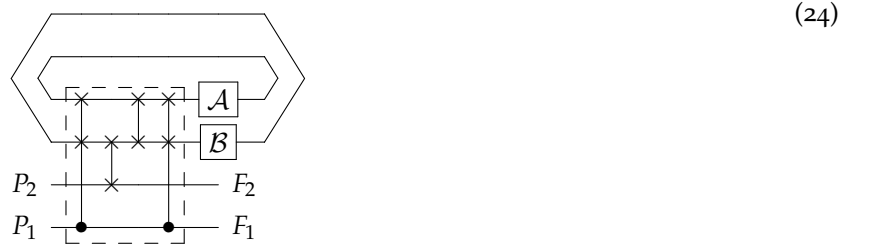
\label{eq:switch2}
\Qcircuit @C=1em @R=0.3em @!R {
\preselection{3}{2.4} & \qw  & \qw & \qw & \qw & \qw & \qw \postselection{3}{2.4} \\
\preselection{1}{0.8} & \qw  & \qw & \qw & \qw & \qw & \qw \postselection{1}{0.8} \\
			      & \qswap     & \qw        & \qswap     & \qswap     & \gate{\mathcal A} & \qw \\
			      & \qswap\qwx & \qswap     & \qswap\qwx & \qswap\qwx & \gate{\mathcal B} & \qw\\
          \lstick{P_2}	      & \qw        & \qswap\qwx & \qw        & \qw        & \rstick{F_2} \qw      &   \\
	  \lstick{P_1}        & \ctrl{-2}  & \qw        & \qw        & \ctrl{-2}  & \rstick{F_1} \qw  \gategroup{3}{2}{6}{5}{.9em}{--} & 
}
\ee
The dashed box singles out the unitary represented by $\ket{w_{\text{switch}2}}$.

An equivalent circuit with P-CTCs for $\ket{w_{\text{switch}2}}$ was obtained in Ref.~\cite{chiribella09}, using a different strategy. We show in Appendix \ref{sec:nswitch} how to construct acausal circuits for $\ket{w_{\text{switch}n}}$ for all $n$. This family of acausal circuits can efficiently create superpositions of $n$ causal orders, by using $n$ P-CTCs, and thereby solve the computational problem proposed in Ref.~\cite{araujo14} using only $n$ queries to the oracles $\mathcal A, \mathcal B$, \ldots, whereas a causally ordered circuit is conjectured to require $\Omega(n^2)$ queries.

\section{An acausal process with no complexity advantage}

The quantum switch cannot violate any causal inequalities \cite{araujo15witnessing}, so it is not the most extreme example of indefinite causal order. One might expect, then, that processes which do violate causal inequalities provide a larger computational advantage. This is however not necessarily the case, as we shall show here by an explicit counterexample: a family of $n$-partite processes which do violate causal inequalities, but can nevertheless be simulated by a causally ordered circuit with only a constant overhead.

To introduce it, we shall first present the tripartite version, show how to simulate it with a causally ordered circuit, and then generalise it to $n$ parties, which we will then show to violate causal inequalities. For $n=3$ the process is a purification of the classical deterministic process $W_\text{det}$ described in Refs.~\cite{baumeler16,araujo16}. In its turn, $W_\text{det}$ is defined as a map from $A_O,B_O,C_O$ to $A_I,B_I,C_I$ that incoherently maps the basis states $\ket{x_0,x_1,x_2}$ to $\ket{f(x_0,x_1,x_2)}$, where $x_i$ are bits and
\be 
f(x_0,x_1,x_2) = (x_2 \land \lnot x_1, x_0 \land \lnot x_2, x_1 \land \lnot x_0).
\ee
Its purification is obtained by making the function $f$ reversible in the standard way: $\ket{x} \mapsto \ket{f(x)}$ becomes $\ket{x}\ket{y} \mapsto \ket{x}\ket{y\oplus f(x)}$. The process vector of this purification\footnote{To check that $\ket{w_{\text{det}3}}$ is indeed a purification of $W_\text{det}$, one needs to check that it is a pure process and that one can recover $W_\text{det}$ from it by inputting the state $\ket{0,0,0}$ in the subsystem $P$ and tracing out the subsystem $F$. It is straightforward to do the latter check. To do the former, we invoke Theorem 2 from \cite{araujo16} and combine it with the proof of validity of $\ket{w_{\text{det}n}}$ shown in Appendix \ref{sec:proof_validity}.} is then
\be\label{eq:wdetpurification3}
\ket{w_{\text{det}3}} = \sum_{\substack{x_0x_1x_2\\y_0y_1y_2}}\ket{y_0,y_1,y_2}\ket{x_0,x_1,x_2}\ket{y_0\oplus x_2 \land \lnot x_1, y_1 \oplus x_0 \land \lnot x_2, y_2\oplus x_1 \land \lnot x_0} \ket{x_0,x_1,x_2},
\ee
where the subsystems are written in the order $P, A_O, B_O, C_O, A_I, B_I, C_I, F$. Rewriting $\ket{w_{\text{det}3}}$ as a quantum circuit with P-CTCs, we get
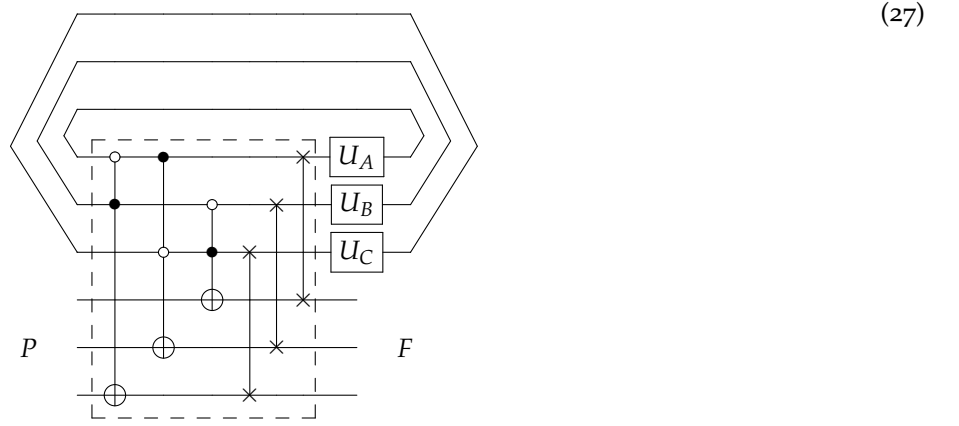
\be\label{eq:wdetacausalcircuit}
\Qcircuit @C=1em @R=0.3em @!R {
\preselection{5}{4.0} & \qw  & \qw & \qw &\qw & \qw & \qw & \qw& \qw\postselection{5}{4.0} \\
\preselection{3}{2.4} & \qw  & \qw & \qw &\qw & \qw & \qw & \qw& \qw\postselection{3}{2.4} \\
\preselection{1}{0.8} & \qw  & \qw & \qw &\qw & \qw & \qw & \qw& \qw\postselection{1}{0.8} \\
			      & \ctrlo{1} & \ctrl{2}  & \qw       & \qw        & \qw        & \qswap     & \gate{U_A} & \qw \\
			      & \ctrl{4}  & \qw       & \ctrlo{1} & \qw        & \qswap     & \qw\qwx    & \gate{U_B} & \qw\\
			      & \qw       & \ctrlo{2} & \ctrl{1}  & \qswap     & \qw\qwx    & \qw\qwx    & \gate{U_C} & \qw\\
			      & \qw       & \qw       & \targ     & \qw\qwx    & \qw\qwx    & \qswap\qwx & \qw &  \\
		\lstick{P\quad}    & \qw       & \targ     & \qw       & \qw\qwx    & \qswap\qwx & \qw        & \rstick{\quad F} \qw &  \\
			      & \targ     & \qw       & \qw       & \qswap\qwx & \qw        & \qw        & \qw \gategroup{4}{2}{9}{7}{.9em}{--} & 
}
\ee

To calculate which transformation from $P$ to $F$ this circuit implements when input with some unitaries $U_A$, $U_B$, and $U_C$, we first rewrite the process as 
\be
\ket{w_{\text{det}3}} = \sum_{\substack{x_0x_1x_2\\y_0y_1y_2}}\ket{y_0,y_1,y_2}\ket{x_0,x_1,x_2}U_{f(x_0,x_1,x_2)}\ket{y_0,y_1,y_2}\ket{x_0,x_1,x_2},
\ee
using the shorthand 
\be
U_{f(x_0,x_1,x_2)} = X^{x_2 \land \lnot x_1} \otimes X^{x_0 \land \lnot x_2} \otimes X^{x_1 \land \lnot x_0},
\ee
where $X$ is the Pauli $X$ matrix.

We then apply $U_A$, $U_B$, and $U_C$ to the systems $A_O,B_O$, and $C_O$, and connect $A_O,B_O,C_O$ to $A_I,B_I,C_I$ using P-CTCs, as per definition \eqref{eq:w_ctc_circuit}. The CJ representation of the unitary $U_G$ that maps $P$ to $F$ is then 
\begin{align}
\Ket{U_G} &= \sum_{\substack{x_0x_1x_2\\y_0y_1y_2}}\ket{y_0,y_1,y_2} \bra{x_0,x_1,x_2} (U_A \otimes U_B \otimes U_C) U_{f(x_0,x_1,x_2)} \ket{y_0,y_1,y_2} \ket{x_0,x_1,x_2}\\
		&= \sum_{x_0x_1x_2} U_{f(x_0,x_1,x_2)} (U_A \otimes U_B \otimes U_C)^T \ket{x_0,x_1,x_2} \ket{x_0,x_1,x_2}. \label{eq:wdet_evolution}
\end{align}
Taking the inverse CJ map of $\Ket{U_G}$ shows us that this unitary does the transformation
\be\label{eq:transformation_det3}
U_{f(x_0,x_1,x_2)} (U_A \otimes U_B \otimes U_C)^\dagger \ket{x_0,x_1,x_2} \mapsto \ket{x_0,x_1,x_2}
\ee
for every $(x_0,x_1,x_2)$. To write this transformation as a circuit from $P$ to $F$, let $U_G'$ be such that $U_G = U_G'(U_A \otimes U_B \otimes U_C)$. Its action can be written in the more convenient way
\begin{align}
|i i i\rangle &\mapsto |i i i \rangle \\
|i 0 1 \rangle & \mapsto(U_A X U_A^\dagger \otimes \mathbb{I}\otimes \mathbb{I}) |i 0 1 \rangle \\
|1 i 0 \rangle  & \mapsto (\mathbb{I} \otimes U_B X U_B^\dagger \otimes \mathbb{I}) |1 i 0 \rangle \\
|0 1 i \rangle  & \mapsto (\mathbb{I}\otimes \mathbb{I} \otimes U_C X U_C^\dagger) |0 1 i \rangle,
\end{align}
where $i \in \{0,1\}$.

This shows that after the application of $U_A \otimes U_B \otimes U_C$, the bits which determine the value of $f(x_0,x_1,x_2)$ become accessible, which makes it easy to implement $U_G'$: we use them to apply $U_{f(x_0,x_1,x_2)}$ to an ancilla, then we apply $(U_A \otimes U_B \otimes U_C)^\dagger$ to the system itself, copy the result of $U_{f(x_0,x_1,x_2)}$ from the ancilla to the system via a CNOT, and apply $U_A \otimes U_B \otimes U_C$ to the system. To finish, we apply $U_{f(x_0,x_1,x_2)}$ a second time to the ancilla, to disentangle it from the system. The end result is a unitary from $P$ to $F$ given by the following circuit:
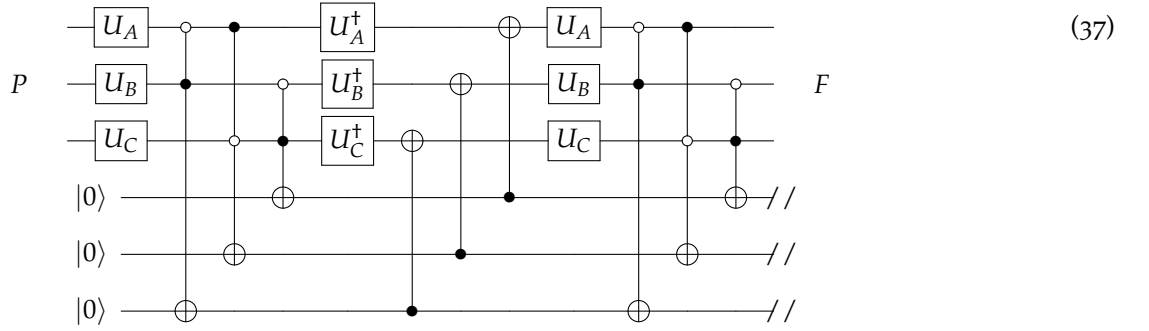
\be\label{eq:wdetcausalcircuit}
\Qcircuit @C=1em @R=0.3em @!R {
& \gate{U_A}         & \ctrlo{1} & \ctrl{2}  & \qw       & \gate{U_A^\dagger} & \qw       & \qw       & \targ     & \gate{U_A} &  \ctrlo{1} & \ctrl{2}  & \qw       & \qw \\
\lstick{P\quad} & \gate{U_B}         & \ctrl{4}  & \qw       & \ctrlo{1} & \gate{U_B^\dagger} & \qw       & \targ     & \qw       & \gate{U_B} &  \ctrl{4}  & \qw       & \ctrlo{1} & \rstick{\quad F} \qw  \\
& \gate{U_C}         & \qw       & \ctrlo{2} & \ctrl{1}  & \gate{U_C^\dagger} & \targ     & \qw       & \qw       & \gate{U_C} &  \qw       & \ctrlo{2} & \ctrl{1}  & \qw \\
& \lstick{\ket{0}} & \qw       & \qw       & \targ     & \qw              & \qw       & \qw       & \ctrl{-3} & \qw      &  \qw       & \qw       & \targ     & \qw {\ \ }{/} {/} & \\
& \lstick{\ket{0}} & \qw       & \targ     & \qw       & \qw              & \qw       & \ctrl{-3} & \qw       & \qw      &  \qw       & \targ     & \qw       & \qw {\ \ }{/} {/} & \\
& \lstick{\ket{0}} & \targ     & \qw       & \qw       & \qw              & \ctrl{-3} & \qw       & \qw       & \qw      &  \targ     & \qw       & \qw       & \qw {\ \ }{/} {/} & \\
}
\ee
To check that this circuit in fact implements the same evolution as in equation \eqref{eq:wdet_evolution}, we first write down its direct CJ representation
\be\label{eq:not_another_label}
\Ket{S} = \sum_{xyz} \ket{x} \ketbra{z}R U_{f(y)} R^\dagger \ketbra{y} R \ket{x}U_{f(z)}U_{f(y)}\ket{0},
\ee
where $R = U_A \otimes U_B \otimes U_C$ and $x,y$, and $z$ are 3-bit strings. Then because
\be\label{eq:mysterious_property}
\bra{z}R U_{f(y)} R^\dagger \ket{y} = 0\mathand \bra{z}R U_{f(z)} R^\dagger \ket{y} = 0
\ee
whenever $f(z) \neq f(y)$, as can easily be checked, we can substitute all occurrences of $f(y)$ with $f(z)$ in equation \eqref{eq:not_another_label}, simplifying it to
\begin{align}
\Ket{S} &= \sum_{xyz} \ket{x} \ketbra{z}R U_{f(z)} R^\dagger \ketbra{y} R \ket{x}U_{f(z)}U_{f(z)}\ket{0} \\
	&= \sum_{xz} \ket{x} \ketbra{z}R U_{f(z)} R^\dagger R \ket{x}\ket{0} \\
	&= \sum_{xz} \ket{x}\bra{x}U_{f(z)}R^T \ket{z} \ket{z}\ket{0} \\
	&= \sum_{z} U_{f(z)}R^T \ket{z} \ket{z}\ket{0},
\end{align}
which is the same unitary as in equation \eqref{eq:wdet_evolution}.

It might be interesting to notice that unlike the quantum switch, we could not find a simulation of the circuit \eqref{eq:wdetacausalcircuit} that does not require the gates $U_A^\dagger$, $U_B^\dagger$, and $U_C^\dagger$. While it might be an interesting puzzle to prove whether this is possible, this question is not relevant for determining the complexity of simulating \eqref{eq:wdetacausalcircuit} and its generalisations, as the complexity of implementing $U^\dagger$ is the same as the complexity of implementing $U$, provided that one knows its decomposition into elementary gates. One might, however, be worried about how to do the causally ordered simulation of \eqref{eq:wdetacausalcircuit} when $U_A$, $U_B$, and $U_C$ are replaced by general CPTP maps, as they will not in general have a positive inverse. This is not a problem, as one only needs to use a purification of the CPTP maps and a simple modification of circuit \eqref{eq:wdetcausalcircuit}, as shown in Appendix \ref{sec:impure_circuit}.

The generalised $n$-partite process is defined analogously to the tripartite one, via an extension of the function $f$ to several parties. The $k$th bit of the extended function given by
\be\label{eq:extended_f}
f(x)_k = x_{k\ominus1} \land \de{\bigwedge_{l=1}^{n-2} \lnot x_{k\oplus l} },
\ee
where addition is done modulo $n$. The purified $n$-partite process is then
\be\label{eq:wdetpurificationn}
\ket{w_{\text{det}n}} = \sum_{xy}\ket{y}\ket{x}U_{f(x)}\ket{y} \ket{x},
\ee
where as before
\be
U_{f(x)} = \bigotimes_{k=0}^{n-1} X^{f(x)_k}
\ee
Unlike for $\ket{w_{\text{det}3}}$, one cannot check the validity of $\ket{w_{\text{det}n}}$ (i.e., that it respects definition \eqref{eq:definition_w}) simply by calculating its trace with a finite number of matrices, so we provide a proof of its validity in Appendix \ref{sec:proof_validity}.

If we represent the reversible version of $f$ by \raisebox{0.5em}{$\Qcircuit @C=0.3em @R=0.1em { & \multigate{1}{f} & \qw \\ & \ghost{f} & \qw }$}\ , the acausal circuit of process \eqref{eq:wdetpurificationn} is 
\be
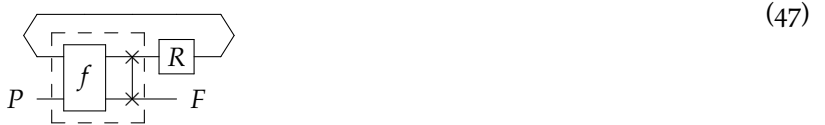
\Qcircuit @C=1em @R=0.3em @!R {
\preselection{1}{0.8} & \qw  & \qw & \qw   & \qw \postselection{1}{0.8} \\
			      & \multigate{1}{f}                 & \qswap     & \gate{R} & \qw \\
	\lstick{P}	      & \ghost{f}             & \qswap\qwx & \rstick{F} \qw \gategroup{2}{2}{3}{3}{.9em}{--} &  \\
}
\ee
where every line represents $n$ qubits, and $R = U_A \otimes U_B \otimes U_C \otimes \ldots$. The circuit that does the causally ordered simulation is analogous to circuit \eqref{eq:wdetcausalcircuit}:
\be
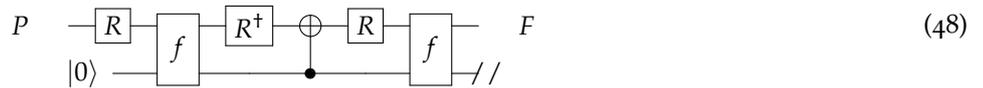
\label{eq:wdetcausalcircuitgeneral}
\Qcircuit @C=1em @R=0.3em @!R {
\lstick{P\quad} & \gate{R}         & \multigate{1}{f} & \gate{R^\dagger} & \targ     & \gate{R} &  \multigate{1}{f} &  \rstick{\quad F} \qw \\
& \lstick{\ket{0}} & \ghost{f}        & \qw              & \ctrl{-1} & \qw      &  \ghost{f}        & \qw {\ \ }{/} {/} & \\
}
\ee
and the proof that it works, even for general CPTP maps, is given in Appendix \ref{sec:impure_circuit}. From this circuit we see that only $3n$ queries to the local operations are needed to simulate $\ket{w_{\text{det}n}}$ in a causally ordered way, proving that this process provides no computational advantage whatsoever.

Now we shall check that $\ket{w_{\text{det}n}}$ does in fact violate causal inequalities for all $n \ge 3$. The causal game we shall consider asks each party $k$ to give output $a_k = 1$ if the input of party $k\ominus1$ is $x_{k\ominus1}=1$ and the input of party $k\oplus1$ is $x_{k\oplus1} = 0$, and to give output $a_k=0$ otherwise. The inputs of the game are promised to be in the set of strings $S$ consisting of the $2n$ translations of the strings $1000\ldots$ and $1100\ldots$. Note that $S$ is the set of strings $x$ for which $f(x) \neq 0$. The probability of success is then given by
\be
p_\text{succ} = \frac1{2n}\sum_{x \in S} p\de{\bigwedge_{k=0}^{n-1} a_k = x_{k\ominus 1} \land \lnot x_{k\oplus 1}}
\ee
Note that if the parties set their output $a_k$ equal to the input given by the process $i_k$ and set their output to the process $o_k$ equal to the input given by the game $x_k$ they win with probability $1$. If the parties are restricted to causally ordered strategies their probability of success is upperbounded by $1-1/n$. Since the game is translation-invariant, we can without loss of generality consider that party $0$ is in the past of everyone else. Since she has no information, she must guess. Since she must output $1$ for only two of the input strings (those for which $(x_{n-1},x_0,x_1)$ is $(100)$ or $(110)$), she guesses $0$, which is correct with probability $1-1/n$. For $n\ge 4$ all the other parties can succeed with probability $1$, as the restriction in the input set means that $x_{k\ominus1}=1$ implies $x_{k\ominus1}=0$, so they simply need to act in the order $k$ before $k + 1$. For $n=3$ this implication is not true, because e.g. $(x_0,x_1,x_2) = (101)$ is a valid input. In this case the probability that the party that comes second in the causal order guesses correctly is less than $1$, and therefore the maximal probability of success is going to be less than $1-1/n$. Nevertheless, it is still valid as an upper bound, so the causal inequality is violated for $n \ge 3$.

\section{Conclusion}

We have shown that process matrices are equivalent to linear post-selected closed timelike curves (P-CTCs), and motivated by this equivalence defined a complexity class -- \bqplctc{} -- as the class of problems that can be efficiently solved by a quantum computer with access to linear P-CTCs. This allowed us to show that \bqplctc is contained in PP, and that a causally ordered quantum circuit with a number of gates exponential in the number of P-CTCs is capable of simulating any process matrix. We expect these upper bounds to be rather loose, however, as the infinite families of process matrices that we do know -- $\ket{w_{\text{switch}n}}$ and $\ket{w_{\text{det}n}}$ -- can be simulated with a polynomial number of gates. We have also found that $\ket{w_{\text{det}n}}$ even violates causal inequalities for all $n$, showing that indefiniteness in the causal order is not \textit{per se} hard to simulate.

\section{Acknowledgements}

We thank Niel de Beaudrap, Časlav Brukner, Fabio Costa, Adrien Feix, David Gross, Tomoyuki Morimae, and Stefan Wolf for useful discussions. We acknowledge support from the Swiss National Science Foundation (SNF), the Austrian Science Fund (FWF) through the Special Research Programme FoQuS, the Doctoral Programme CoQuS and the projects No. P-24621 and I-2526. This work has been supported by the Excellence Initiative of the German Federal and State Governments (Grant ZUK 81). P. A. G. is also supported by FQRNT (Québec).

\bibliographystyle{linksen}
\bibliography{biblio}

\providecommand{\href}[2]{#2}\begingroup\raggedright\begin{thebibliography}{10}

\bibitem{godel49}
K.~G\"odel, ``An Example of a New Type of Cosmological Solutions of Einstein's
  Field Equations of Gravitation,''
  \href{http://dx.doi.org/10.1103/RevModPhys.21.447}{{\em Rev. Mod. Phys.}
  {\bfseries 21}, 447--450 (1949)}.

\bibitem{lanczos24}
K.~Lanczos, ``Über eine stationäre Kosmologie im Sinne der Einsteinschen
  Gravitationstheorie,'' \href{http://dx.doi.org/10.1007/BF01328251}{{\em
  Zeitschrift für Physik} {\bfseries 21}, (1924)}.

\bibitem{morris88}
M.~S. Morris, K.~S. Thorne, and U.~Yurtsever, ``Wormholes, Time Machines, and
  the Weak Energy Condition,''
  \href{http://dx.doi.org/10.1103/PhysRevLett.61.1446}{{\em Phys. Rev. Lett.}
  {\bfseries 61}, 1446--1449 (1988)}.

\bibitem{novikov89}
I.~D. Novikov, ``An analysis of the operation of a time machine,'' {\em Soviet
  physics, JETP} {\bfseries 68}, 439 (1989).
  \url{http://www.jetp.ac.ru/cgi-bin/dn/e_068_03_0439.pdf}.

\bibitem{hawking92}
S.~W. Hawking, ``Chronology protection conjecture,''
  \href{http://dx.doi.org/10.1103/PhysRevD.46.603}{{\em Phys. Rev. D}
  {\bfseries 46}, 603--611 (1992)}.

\bibitem{deutsch91}
D.~{Deutsch}, ``{Quantum mechanics near closed timelike lines},''
  \href{http://dx.doi.org/10.1103/PhysRevD.44.3197}{{\em Phys. Rev.~D}
  {\bfseries 44}, 3197--3217 (1991)}.

\bibitem{brun09}
T.~A. {Brun}, J.~{Harrington}, and M.~M. {Wilde}, ``{Localized Closed Timelike
  Curves Can Perfectly Distinguish Quantum States},''
  \href{http://dx.doi.org/10.1103/PhysRevLett.102.210402}{{\em Phys. Rev.
  Lett.} {\bfseries 102}, 210402 (2009)},
  \href{http://arxiv.org/abs/0811.1209}{{\ttfamily arXiv:0811.1209
  [quant-ph]}}.

\bibitem{ahn13}
D.~{Ahn}, C.~R. {Myers}, T.~C. {Ralph}, and R.~B. {Mann}, ``{Quantum-state
  cloning in the presence of a closed timelike curve},''
  \href{http://dx.doi.org/10.1103/PhysRevA.88.022332}{{\em Phys. Rev.~A}
  {\bfseries 88}, 022332 (2013)},
  \href{http://arxiv.org/abs/1207.6062}{{\ttfamily arXiv:1207.6062
  [quant-ph]}}.

\bibitem{brun13}
T.~A. {Brun}, M.~M. {Wilde}, and A.~{Winter}, ``{Quantum State Cloning Using
  Deutschian Closed Timelike Curves},''
  \href{http://dx.doi.org/10.1103/PhysRevLett.111.190401}{{\em Phys. Rev.
  Lett.} {\bfseries 111}, 190401 (2013)},
  \href{http://arxiv.org/abs/1306.1795}{{\ttfamily arXiv:1306.1795
  [quant-ph]}}.

\bibitem{pienaar13}
J.~L. {Pienaar}, T.~C. {Ralph}, and C.~R. {Myers}, ``{Open Timelike Curves
  Violate Heisenberg's Uncertainty Principle},''
  \href{http://dx.doi.org/10.1103/PhysRevLett.110.060501}{{\em Phys. Rev.
  Lett.} {\bfseries 110}, 060501 (2013)},
  \href{http://arxiv.org/abs/1206.5485}{{\ttfamily arXiv:1206.5485
  [quant-ph]}}.

\bibitem{aaronson09}
S.~{Aaronson} and J.~{Watrous}, ``{Closed timelike curves make quantum and
  classical computing equivalent},''
  \href{http://dx.doi.org/10.1098/rspa.2008.0350}{{\em Proc. R. Soc. Lond. A}
  {\bfseries 465}, 631--647 (2009)},
  \href{http://arxiv.org/abs/0808.2669}{{\ttfamily arXiv:0808.2669
  [quant-ph]}}.

\bibitem{politzer94}
H.~D. Politzer, ``Path integrals, density matrices, and information flow with
  closed timelike curves,''
  \href{http://dx.doi.org/10.1103/PhysRevD.49.3981}{{\em Phys. Rev. D}
  {\bfseries 49}, 3981--3989 (1994)}.

\bibitem{bennett05}
C.~H. Bennett and B.~Schumacher, ``Teleportation, Simulated Time Travel, and
  How to Flirt with Someone Who Has Fallen into a Black Hole,'' 2005.
\newblock
  \url{http://web.archive.org/web/20060207123032/http://www.research.ibm.com/people/b/bennetc/QUPONBshort.pdf}.
  Talk in QUPON.

\bibitem{svetlichny11}
G.~{Svetlichny}, ``{Time Travel: Deutsch vs. Teleportation},''
  \href{http://dx.doi.org/10.1007/s10773-011-0973-x}{{\em Int. J. Theor. Phys.}
  {\bfseries 50}, 3903--3914 (2011)},
  \href{http://arxiv.org/abs/0902.4898}{{\ttfamily arXiv:0902.4898
  [quant-ph]}}.

\bibitem{brun12}
T.~A. {Brun} and M.~M. {Wilde}, ``{Perfect State Distinguishability and
  Computational Speedups with Postselected Closed Timelike Curves},''
  \href{http://dx.doi.org/10.1007/s10701-011-9601-0}{{\em Found. Phys.}
  {\bfseries 42}, 341--361 (2012)},
  \href{http://arxiv.org/abs/1008.0433}{{\ttfamily arXiv:1008.0433
  [quant-ph]}}.

\bibitem{lloyd11}
S.~{Lloyd}, L.~{Maccone}, R.~{Garcia-Patron}, V.~{Giovannetti}, Y.~{Shikano},
  S.~{Pirandola}, L.~A. {Rozema}, A.~{Darabi}, Y.~{Soudagar}, L.~K. {Shalm},
  and A.~M. {Steinberg}, ``{Closed Timelike Curves via Postselection: Theory
  and Experimental Test of Consistency},''
  \href{http://dx.doi.org/10.1103/PhysRevLett.106.040403}{{\em Phys. Rev.
  Lett.} {\bfseries 106}, 040403 (2011)},
  \href{http://arxiv.org/abs/1005.2219}{{\ttfamily arXiv:1005.2219
  [quant-ph]}}.

\bibitem{aaronson05b}
S.~{Aaronson}, ``{NP-complete Problems and Physical Reality},''
  \href{http://dx.doi.org/10.1145/1052796.1052804}{{\em SIGACT News} {\bfseries
  36}, 30--52 (2005)}, \href{http://arxiv.org/abs/quant-ph/0502072}{{\ttfamily
  arXiv:quant-ph/0502072}}.

\bibitem{oreshkov12}
O.~{Oreshkov}, F.~{Costa}, and {\v C}.~{Brukner}, ``{Quantum correlations with
  no causal order},'' \href{http://dx.doi.org/10.1038/ncomms2076}{{\em Nat.
  Commun.} {\bfseries 3}, 1092 (2012)},
  \href{http://arxiv.org/abs/1105.4464}{{\ttfamily arXiv:1105.4464
  [quant-ph]}}.

\bibitem{araujo16}
M.~{Araújo}, A.~{Feix}, M.~{Navascués}, and {\v C}.~{Brukner}, ``{A
  purification postulate for quantum mechanics with indefinite causal order},''
  \href{http://dx.doi.org/10.22331/q-2017-04-26-10}{{\em Quantum} {\bfseries
  1}, 10 (2017)}, \href{http://arxiv.org/abs/1611.08535}{{\ttfamily
  arXiv:1611.08535 [quant-ph]}}.

\bibitem{baumeler13}
{\"A}.~Baumeler and S.~Wolf, ``Perfect signaling among three parties violating
  predefined causal order,''
  \href{http://dx.doi.org/10.1109/ISIT.2014.6874888}{in {\em Information Theory
  (ISIT), 2014 IEEE International Symposium on}}, pp.~526--530.
\newblock 2014.
\newblock \href{http://arxiv.org/abs/1312.5916}{{\ttfamily arXiv:1312.5916
  [quant-ph]}}.

\bibitem{baumeler14}
{\"A}.~Baumeler, A.~Feix, and S.~Wolf, ``{Maximal incompatibility of locally
  classical behavior and global causal order in multi-party scenarios},''
  \href{http://dx.doi.org/10.1103/PhysRevA.90.042106}{{\em Phys. Rev. A}
  {\bfseries 90}, 042106 (2014)},
  \href{http://arxiv.org/abs/1403.7333}{{\ttfamily arXiv:1403.7333
  [quant-ph]}}.

\bibitem{branciard15simplest}
C.~Branciard, M.~Araújo, A.~Feix, F.~Costa, and {\v C}.~Brukner, ``The
  simplest causal inequalities and their violation,''
  \href{http://dx.doi.org/10.1088/1367-2630/18/1/013008}{{\em New J. Phys.}
  {\bfseries 18}, 013008 (2015)},
  \href{http://arxiv.org/abs/1508.01704}{{\ttfamily arXiv:1508.01704
  [quant-ph]}}.

\bibitem{baumeler16}
{\"A}.~{Baumeler} and S.~{Wolf}, ``{The space of logically consistent classical
  processes without causal order},''
  \href{http://dx.doi.org/10.1088/1367-2630/18/1/013036}{{\em New J.~Phys.}
  {\bfseries 18}, 013036 (2016)},
  \href{http://arxiv.org/abs/1507.01714}{{\ttfamily arXiv:1507.01714
  [quant-ph]}}.

\bibitem{oreshkov15}
O.~{Oreshkov} and C.~{Giarmatzi}, ``{Causal and causally separable
  processes},'' \href{http://dx.doi.org/10.1088/1367-2630/18/9/093020}{{\em New
  J. Phys.} {\bfseries 18}, 093020 (2015)},
  \href{http://arxiv.org/abs/1506.05449}{{\ttfamily arXiv:1506.05449
  [quant-ph]}}.

\bibitem{abbott16}
A.~A. {Abbott}, C.~{Giarmatzi}, F.~{Costa}, and C.~{Branciard}, ``{Multipartite
  Causal Correlations: Polytopes and Inequalities},''
  \href{http://dx.doi.org/10.1103/PhysRevA.94.032131}{{\em Phys. Rev.~A}
  {\bfseries 94}, 032131 (2016)},
  \href{http://arxiv.org/abs/1608.01528}{{\ttfamily arXiv:1608.01528
  [quant-ph]}}.

\bibitem{chiribella09}
G.~{Chiribella}, G.~M. {D'Ariano}, P.~{Perinotti}, and B.~{Valiron}, ``{Quantum
  computations without definite causal structure},''
  \href{http://dx.doi.org/10.1103/PhysRevA.88.022318}{{\em Phys. Rev.~A}
  {\bfseries 88}, 022318 (2013)},
  \href{http://arxiv.org/abs/0912.0195}{{\ttfamily arXiv:0912.0195
  [quant-ph]}}.

\bibitem{chiribella12}
G.~{Chiribella}, ``{Perfect discrimination of no-signalling channels via
  quantum superposition of causal structures},''
  \href{http://dx.doi.org/10.1103/PhysRevA.86.040301}{{\em Phys. Rev.~A}
  {\bfseries 86}, 040301 (2012)},
  \href{http://arxiv.org/abs/1109.5154}{{\ttfamily arXiv:1109.5154
  [quant-ph]}}.

\bibitem{araujo14}
M.~{Araújo}, F.~{Costa}, and {\v C}.~{Brukner}, ``{Computational Advantage
  from Quantum-Controlled Ordering of Gates},''
  \href{http://dx.doi.org/10.1103/PhysRevLett.113.250402}{{\em Phys. Rev.
  Lett.} {\bfseries 113}, 250402 (2014)},
  \href{http://arxiv.org/abs/1401.8127}{{\ttfamily arXiv:1401.8127
  [quant-ph]}}.

\bibitem{baumeler16c}
{\"A}.~{Baumeler} and S.~{Wolf}, ``{Device-independent test of causal order and
  relations to fixed-points},''
  \href{http://dx.doi.org/10.1088/1367-2630/18/3/035014}{{\em New J.~Phys.}
  {\bfseries 18}, 035014 (2016)},
  \href{http://arxiv.org/abs/1511.05444}{{\ttfamily arXiv:1511.05444
  [quant-ph]}}.

\bibitem{baumeler16d}
{\"A}.~{Baumeler} and S.~{Wolf}, ``{Non-causal computation},''
  \href{http://dx.doi.org/10.3390/e19070326}{{\em Entropy} {\bfseries 19}, 326
  (2017)}, \href{http://arxiv.org/abs/1601.06522}{{\ttfamily arXiv:1601.06522
  [quant-ph]}}.

\bibitem{baumeler16e}
{\"A}.~{Baumeler} and S.~{Wolf}, ``{Computational tameness of classical
  non-causal models},'' \href{http://arxiv.org/abs/1611.05641}{{\ttfamily
  arXiv:1611.05641 [quant-ph]}}.

\bibitem{procopio_experimental_2014}
L.~M. Procopio, A.~Moqanaki, M.~Araújo, F.~Costa, I.~A. Calafell, E.~G. Dowd,
  D.~R. Hamel, L.~A. Rozema, {\v C}.~Brukner, and P.~Walther, ``Experimental
  superposition of orders of quantum gates,''
  \href{http://dx.doi.org/10.1038/ncomms8913}{{\em Nat. Commun.} {\bfseries 6},
  7913 (2015)}, \href{http://arxiv.org/abs/1412.4006}{{\ttfamily
  arXiv:1412.4006 [quant-ph]}}.

\bibitem{rubino16}
G.~{Rubino}, L.~A. {Rozema}, A.~{Feix}, M.~{Ara{\'u}jo}, J.~M. {Zeuner}, L.~M.
  {Procopio}, {\v C}.~{Brukner}, and P.~{Walther}, ``Experimental verification
  of an indefinite causal order,''
  \href{http://dx.doi.org/10.1126/sciadv.1602589}{{\em Sci. Adv.} {\bfseries
  3}, 1602589 (2017)}, \href{http://arxiv.org/abs/1608.01683}{{\ttfamily
  arXiv:1608.01683 [quant-ph]}}.

\bibitem{baumeler17}
{\"A}.~Baumeler, F.~Costa, T.~Ralph, S.~Wolf, and M.~Zych. (in preparation).

\bibitem{araujo15witnessing}
M.~Araújo, C.~Branciard, F.~Costa, A.~Feix, C.~Giarmatzi, and {\v C}.~Brukner,
  ``Witnessing causal nonseparability,''
  \href{http://dx.doi.org/10.1088/1367-2630/17/10/102001}{{\em New J. Phys.}
  {\bfseries 17}, 102001 (2015)},
  \href{http://arxiv.org/abs/1506.03776}{{\ttfamily arXiv:1506.03776
  [quant-ph]}}.

\bibitem{dias11}
R.~{Dias da Silva}, E.~F. {Galv{\~a}o}, and E.~{Kashefi}, ``{Closed timelike
  curves in measurement-based quantum computation},''
  \href{http://dx.doi.org/10.1103/PhysRevA.83.012316}{{\em Phys. Rev.~A}
  {\bfseries 83}, 012316 (2011)},
  \href{http://arxiv.org/abs/1003.4971}{{\ttfamily arXiv:1003.4971
  [quant-ph]}}.

\bibitem{oreshkov16}
O.~{Oreshkov} and N.~{Cerf}, ``{Operational quantum theory without predefined
  time},'' \href{http://dx.doi.org/10.1088/1367-2630/18/7/073037}{{\em New
  J.~Phys.} {\bfseries 18}, 073037 (2016)},
  \href{http://arxiv.org/abs/1406.3829}{{\ttfamily arXiv:1406.3829
  [quant-ph]}}.

\bibitem{costa14private}
F.~Costa. Private communication (2014).

\bibitem{silva17}
R.~{Silva}, Y.~{Guryanova}, A.~J. {Short}, P.~{Skrzypczyk}, N.~{Brunner}, and
  S.~{Popescu}, ``{Connecting processes with indefinite causal order and
  multi-time quantum states},''
  \href{http://dx.doi.org/10.1088/1367-2630/aa84fe}{{\em New J.~Phys.}
  {\bfseries 19}, 103022 (2017)},
  \href{http://arxiv.org/abs/1701.08638}{{\ttfamily arXiv:1701.08638
  [quant-ph]}}.

\bibitem{chiribella09b}
G.~{Chiribella}, G.~M. {D'Ariano}, and P.~{Perinotti}, ``{Theoretical framework
  for quantum networks},''
  \href{http://dx.doi.org/10.1103/PhysRevA.80.022339}{{\em Phys. Rev.~A}
  {\bfseries 80}, 022339 (2009)},
  \href{http://arxiv.org/abs/0904.4483}{{\ttfamily arXiv:0904.4483
  [quant-ph]}}.

\bibitem{aaronson05}
S.~{Aaronson}, ``{Quantum Computing, Postselection, and Probabilistic
  Polynomial-Time},'' \href{http://dx.doi.org/10.1098/rspa.2005.1546}{{\em
  Proc. R. Soc. A} {\bfseries 461}, 3473--3482 (2005)},
  \href{http://arxiv.org/abs/quant-ph/0412187}{{\ttfamily
  arXiv:quant-ph/0412187}}.

\bibitem{arora09}
S.~Arora and B.~Barak, {\em Computational Complexity: A Modern Approach}.
\newblock Cambridge University Press, 2009.

\bibitem{aaronson14}
S.~Aaronson, ``PostBQP Postscripts: A Confession of Mathematical Errors,''
  2014.
\newblock \url{http://www.scottaaronson.com/blog/?p=2072}.

\bibitem{morimae15}
T.~{Morimae} and H.~{Nishimura}, ``{Quantum interpretations of AWPP and APP},''
  {\em Quant. Inf. and Comp.} {\bfseries 16}, 0498--0514 (2016),
  \href{http://arxiv.org/abs/1502.00067}{{\ttfamily arXiv:1502.00067
  [quant-ph]}}.

\bibitem{wikifactoradic}
Wikipedia, ``Factorial number system.''
\newblock \url{https://en.wikipedia.org/wiki/Factorial_number_system}.

\end{thebibliography}\endgroup

\appendix 

\section{\texorpdfstring{$n$-partite acausal circuit for the quantum switch}{n-partite acausal circuit for the quantum switch}}\label{sec:nswitch}

The generalisation to $n$ parties is based on the scheme described in Appendix E of Ref.~\cite{araujo14}. It applies the permutation of $n$ parties labelled by $s \in [0,n!-1]$ to the target system $\ket{\psi}$ when the control system $\ket{C}$ is in the state $s$. The encoding of $s$ into qubits is, however, a bit elaborated. To do it, first write $s$ in the factoradic basis \cite{wikifactoradic}, \ie, find the coefficients $(a_{n-1},\ldots,a_1)$ such that $a_k \le k$ and 
\be s = \sum_{k=1}^{n-1} a_k k! \ee
Now take each factoradic digit $a_k$ and rewrite it in the unary basis, \ie, define the coefficients $(b_{k,k},b_{k,k-1},\ldots,b_{k,1})$ to be $b_{k,i} = 1$ for $1 \le i \le a_k$ and $b_{k,i} = 0$ for $a_k + 1 \le i \le k$. For example, if $n=4$ and $s = 13$ the coefficients $(a_3,a_2,a_1)$ are $(2,0,1)$ which, when written in unary become $(b_{33},b_{32},b_{31}) = (0,1,1)$, $(b_{22},b_{21}) = (0,0)$, and $(b_{11}) = (1)$. The control system is defined then as $\ket{C} = \bigotimes_{i=1}^{n-1}\bigotimes_{j=1}^{i} \ket{b_{ij}}$. 

The crucial part of the circuit is then a series of controlled SWAPs: a SWAP between the wires 1 and 2 controlled on $\ket{b_{11}}$, followed by a SWAP between wires 2 and 3 controlled on $\ket{b_{21}}$, followed by a SWAP between wires 1 and 2 controlled on $\ket{b_{22}}$, followed by a SWAP between wires 3 and 4 controlled on $\ket{b_{31}}$, followed by a SWAP between wires 2 and 3 controlled on $\ket{b_{32}}$, followed by a SWAP between wires 1 and 2 controlled on $\ket{b_{31}}$, and so on. It is important to notice that although this encoding scheme will not produce all possible\footnote{Which should be obvious if you notice that it uses $n(n-1)/2$ qubits to encode a number between $0$ and $n!-1$ instead of the minimal $\lceil \log_2 n! \rceil$ qubits.} bit strings in $\ket{C}$ (for example the bit string $(b_{22},b_{21}) = (1,0)$ will never appear), the ``forbidden'' bit strings do not cause any problems. If one inputs one of them then the circuit will just produce a valid permutation, that however could be produced anyway using a ``valid'' input.

From this encoding it is then straightforward to write down the acausal circuit. For $n=2$ parties it is shown in equation \eqref{eq:switch2} in the main text. For $n=3$ the explicit circuit is
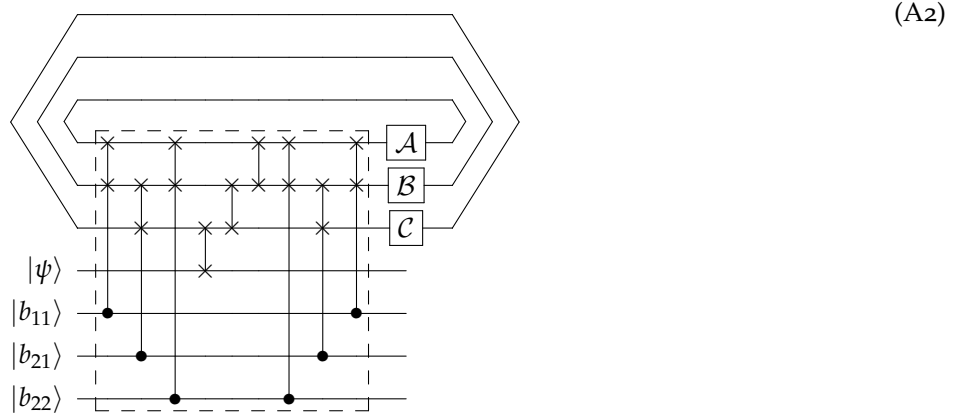
\be
\Qcircuit @C=1em @R=0.3em @!R {
\preselection{5}{4.0} & \qw  & \qw & \qw & \qw & \qw & \qw & \qw & \qw & \qw & \qw & \qw\postselection{5}{4.0} \\
\preselection{3}{2.4} & \qw  & \qw & \qw & \qw & \qw & \qw & \qw & \qw & \qw & \qw & \qw\postselection{3}{2.4} \\
\preselection{1}{0.8} & \qw  & \qw & \qw & \qw & \qw & \qw & \qw & \qw & \qw & \qw & \qw\postselection{1}{0.8} \\
			      & \qswap     & \qw        & \qswap     & \qw       & \qw       & \qswap    & \qswap     & \qw        & \qswap     & \gate{\mathcal A} & \qw \\
			      & \qswap\qwx & \qswap     & \qswap\qwx & \qw       & \qswap    & \qswap\qwx& \qswap\qwx & \qswap     & \qswap\qwx & \gate{\mathcal B} & \qw \\
			      & \qw        & \qswap\qwx & \qw        & \qswap    & \qswap\qwx& \qw       & \qw        & \qswap\qwx & \qw        & \gate{\mathcal C} & \qw \\
   \lstick{\ket{\psi}}        & \qw        & \qw        & \qw        & \qswap\qwx& \qw       & \qw       & \qw        & \qw        & \qw        & \qw      &  \\
   \lstick{\ket{b_{11}}}      & \ctrl{-3}  & \qw        & \qw        & \qw       & \qw       & \qw       & \qw        & \qw        & \ctrl{-3}  & \qw      &  \\
   \lstick{\ket{b_{21}}}      & \qw        & \ctrl{-3}  & \qw        & \qw       & \qw       & \qw       & \qw        & \ctrl{-3}  & \qw        & \qw      &  \\
   \lstick{\ket{b_{22}}}      & \qw        & \qw        & \ctrl{-5}  & \qw       & \qw       & \qw       & \ctrl{-5}  & \qw        & \qw        & \qw \gategroup{4}{2}{10}{10}{.9em}{--} & 
}
\ee
and for $n=4$ the explicit circuit is
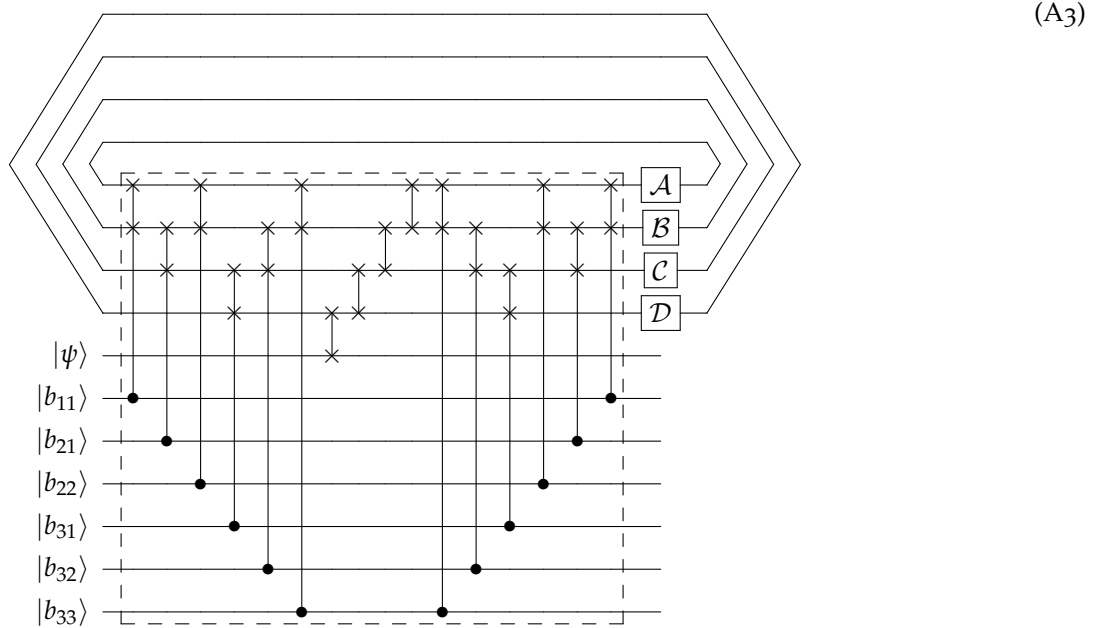
\be
\Qcircuit @C=1em @R=0.3em @!R {
\preselection{7}{5.6} & \qw & \qw & \qw & \qw & \qw & \qw & \qw & \qw & \qw & \qw & \qw & \qw& \qw& \qw& \qw& \qw& \qw& \qw\postselection{7}{5.6} \\
\preselection{5}{4.0} & \qw & \qw & \qw & \qw & \qw & \qw & \qw & \qw & \qw & \qw & \qw & \qw& \qw& \qw& \qw& \qw& \qw& \qw\postselection{5}{4.0} \\
\preselection{3}{2.4} & \qw & \qw & \qw & \qw & \qw & \qw & \qw & \qw & \qw & \qw & \qw & \qw& \qw& \qw& \qw& \qw& \qw& \qw\postselection{3}{2.4} \\
\preselection{1}{0.8} & \qw & \qw & \qw & \qw & \qw & \qw & \qw & \qw & \qw & \qw & \qw & \qw& \qw& \qw& \qw& \qw& \qw& \qw\postselection{1}{0.8} \\
			      & \qswap    & \qw       & \qswap    & \qw       & \qw       & \qswap    & \qw       & \qw       & \qw       & \qswap    & \qswap    & \qw       & \qw        & \qswap     & \qw        & \qswap     & \gate{\mathcal A} & \qw \\
			      & \qswap\qwx& \qswap    & \qswap\qwx& \qw       & \qswap    & \qswap\qwx& \qw       & \qw       & \qswap    & \qswap\qwx& \qswap\qwx& \qswap    & \qw        & \qswap\qwx & \qswap     & \qswap\qwx & \gate{\mathcal B} & \qw \\
			      & \qw       & \qswap\qwx& \qw       & \qswap    & \qswap\qwx& \qw       & \qw       & \qswap    & \qswap\qwx& \qw       & \qw       & \qswap\qwx& \qswap     & \qw        & \qswap\qwx & \qw        & \gate{\mathcal C} & \qw \\
			      & \qw       & \qw       & \qw       & \qswap\qwx& \qw       & \qw       & \qswap    & \qswap\qwx& \qw       & \qw       & \qw       & \qw       & \qswap\qwx & \qw        & \qw        & \qw        & \gate{\mathcal D} & \qw \\      
   \lstick{\ket{\psi}}        & \qw       & \qw       & \qw       & \qw       & \qw       & \qw       & \qswap\qwx& \qw       & \qw       & \qw       & \qw       & \qw       & \qw        & \qw        & \qw        & \qw        & \qw      &  \\
   \lstick{\ket{b_{11}}}      & \ctrl{-4} & \qw       & \qw       & \qw       & \qw       & \qw       & \qw       & \qw       & \qw       & \qw       & \qw       & \qw       & \qw        & \qw        & \qw        & \ctrl{-4}  & \qw      &  \\
   \lstick{\ket{b_{21}}}      & \qw       & \ctrl{-4} & \qw       & \qw       & \qw       & \qw       & \qw       & \qw       & \qw       & \qw       & \qw       & \qw       & \qw        & \qw        & \ctrl{-4}  & \qw        & \qw      &  \\      
   \lstick{\ket{b_{22}}}      & \qw       & \qw       & \ctrl{-6} & \qw       & \qw       & \qw       & \qw       & \qw       & \qw       & \qw       & \qw       & \qw       & \qw        & \ctrl{-6}  & \qw        & \qw        & \qw      &  \\      
   \lstick{\ket{b_{31}}}      & \qw       & \qw       & \qw       & \ctrl{-5} & \qw       & \qw       & \qw       & \qw       & \qw       & \qw       & \qw       & \qw       & \ctrl{-5}  & \qw        & \qw        & \qw        & \qw      &  \\      
   \lstick{\ket{b_{32}}}      & \qw       & \qw       & \qw       & \qw       & \ctrl{-7} & \qw       & \qw       & \qw       & \qw       & \qw       & \qw       & \ctrl{-7} & \qw        & \qw        & \qw        & \qw        & \qw      &  \\      
   \lstick{\ket{b_{33}}}      & \qw       & \qw       & \qw       & \qw       & \qw       & \ctrl{-9} & \qw       & \qw       & \qw       & \qw       & \ctrl{-9} & \qw       & \qw        & \qw        & \qw        & \qw        & \qw \gategroup{5}{2}{15}{17}{.9em}{--} & 
}
\ee

\section{Proof of validity of the multipartite process}\label{sec:proof_validity}

For the process $\ket{w_{\text{det}n}}$ defined in equation \eqref{eq:wdetpurificationn} to be valid, it needs to output a valid CPTP map whenever the parties input a valid CPTP $A^k$, that is the operator 
\be\label{eq:definition_w_n}
G_A = \tr_{A^0_IA^0_O\ldots A^{n-1}_IA^{n-1}_O} \De{\ketbra{w_{\text{det}n}}\de{\bigotimes_{k=0}^{n-1} {A^k}^T}}
\ee
must be a valid CPTP map for all CPTP $A^k$.

Let then $A_i$ be the Kraus operators of the CPTP map $A = \bigotimes_{k=0}^{n-1} {A^k}$, i.e., the operators such that
\be
A = \sum_{a,b,i} \ket{a}\bra{b}\otimes A_i \ket{a}\bra{b} {A_i}^\dagger.
\ee
Then we have also
\be
A_i = \bigotimes_{k=0}^{n-1} A^k_{i_k},
\ee
where $A^k_{i_k}$ are the Kraus operators of the individual CPTP maps $A^k$.

Substituting the definitions of $A$ and $\ket{w_{\text{det}n}}$ in equation \eqref{eq:definition_w_n} we get
\begin{align}
G_A &= \sum_{\substack{i,x,y\\x',y',a,b}} \ket{y}\bra{y'} \tr\De{{A_i^\dagger}^T\ket{b}\bra{a}A_i^T\ket{x}\bra{x'}} \tr\Big[\ket{b}\bra{a}   U_{f(x)}\ket{y} \bra{y'}U_{f(x')}\Big] \ket{x}\bra{x'} \\
    &= \sum_{\substack{i,x,y\\x',y'}} \ket{y}\bra{y'} \otimes \bra{x}A_iU_{f(x)}\ket{y} \bra{y'}U_{f(x')} A_i^\dagger \ket{x'} \otimes \ket{x}\bra{x'} \\
    &= \sum_{\substack{i,x,y\\x',y'}} \ket{y}\bra{y}U_{f(x)}A_i^T\ket{x}\bra{x'} {A_i^\dagger}^T U_{f(x')} \ket{y'}\bra{y'}\otimes \ket{x}\bra{x'} \\
    &= \sum_{i,x,x'} U_{f(x)}A_i^T\ket{x}\bra{x'} {A_i^\dagger}^T U_{f(x')}\otimes \ket{x}\bra{x'}
\end{align}
Since $G_A$ is trivially positive, to check that it is CPTP we only need to check that it is trace preserving, i.e., that
\be
\tr_{F}G_A = \id^P,
\ee
so we need to show that
\be
\sum_{i,x} U_{f(x)}A_i^\dagger \ket{x}\bra{x} {A_i} U_{f(x)} = \id^P,
\ee
where we also applied a transpose to both sides. To do that, first we break down the sum on \emph{lhs} according to the value of $f$:
\be
\sum_{i,x} U_{f(x)}A_i^\dagger \ket{x}\bra{x} {A_i} U_{f(x)} = \sum_{i,x;f(x)=0} A_i^\dagger \ket{x}\bra{x} {A_i} + \sum_k \sum_{i,x;f(x)_k = 1} U_{f(x)}A_i^\dagger \ket{x}\bra{x} {A_i} U_{f(x)}.
\ee
This decomposition is valid because $f$ has the property that if the $k$th bit of $f(x)$ is one, then all other bits of $f(x)$ are equal to zero. Furthermore, there are only two inputs $x$ for which the $k$th bit of $f(x)$ is one: the string where $x_k=0$, $x_{k\ominus1}=1$ and all other $x_i$ are zero, and the string where $x_k=1$, $x_{k\ominus1}=1$ and all other $x_i$ are zero.

Concentrating on the term $k=0$ of the \emph{rhs} to reduce clutter, we have that
\begin{align}
\sum_{i,x;f(x)_0 = 1} U_{f(x)}A_i^\dagger \ket{x}\bra{x} {A_i} U_{f(x)} = & \sum_i X \otimes \id^{\otimes n-1} A_i^\dagger \Big(\ketbra{00^{n-2}1} + \ketbra{10^{n-2}1}\Big) A_i X \otimes \id^{\otimes n-1} \\
   =&\sum_i X \otimes \id^{\otimes n-1} A_i^\dagger \de{ \id \otimes \ketbra{0^{n-2}1} } A_i X \otimes \id^{\otimes n-1} \\
   =&\sum_{i_k;k\neq 0}  \bigotimes_{k=1}^{n-1} {A^k_{i_k}}^\dagger \de{X \Big(\sum_{i_0} A_{i_0}^\dagger A_{i_0}\Big) X\otimes \ketbra{0^{n-2}1} } \bigotimes_{k=1}^{n-1} A^k_{i_k} \\
   =&\sum_{i_k;k\neq 0}  \bigotimes_{k=1}^{n-1} {A^k_{i_k}}^\dagger \de{\id \otimes \ketbra{0^{n-2}1} } \bigotimes_{k=1}^{n-1} A^k_{i_k} \label{eq:krausidentity}\\
   =& \sum_i A_i^\dagger \Big(\ketbra{00^{n-2}1} + \ketbra{10^{n-2}1}\Big) A_i \\
   =& \sum_{i,x;f(x)_0 = 1} A_i^\dagger \ket{x}\bra{x} A_i,
\end{align}
where $0^{n-2}$ is the string composed of $n-2$ zeros, and line \eqref{eq:krausidentity} follows from the previous one because by assumption the map $A^0$ is CPTP, which in the Kraus representation means that $\sum_{i_0} {A^0_{i_0}}^\dagger A^0_{i_0} = \id$. 

Since the analogous reasoning is valid for all parties $k$, we have that 
\be
\sum_{i,x;f(x)_k = 1} U_{f(x)}A_i^\dagger \ket{x}\bra{x} {A_i} U_{f(x)} = \sum_{i,x;f(x)_k = 1} A_i^\dagger \ket{x}\bra{x} A_i
\ee
and therefore
\be
\sum_{i,x} U_{f(x)}A_i^\dagger \ket{x}\bra{x} {A_i} U_{f(x)} = \sum_{i,x} A_i^\dagger \ket{x}\bra{x} {A_i} = \id^P.
\ee
\qed

\section{Ordered simulation}\label{sec:impure_circuit}

In this Appendix we will show how to simulate with a causally ordered circuit the same evolution from $P$ to $F$ as the one induced by $\ket{w_{\text{det}n}}$ when the parties apply general CPTP maps, that is, the evolution induced by the circuit
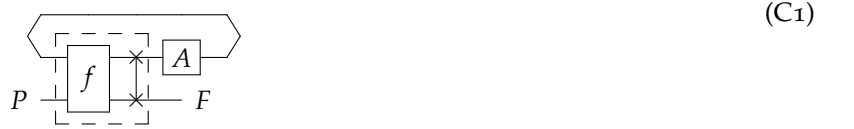
\be\label{eq:circuit_detn_cptp}
\Qcircuit @C=1em @R=0.3em @!R {
\preselection{1}{0.8} & \qw  & \qw & \qw   & \qw \postselection{1}{0.8} \\
			      & \multigate{1}{f}                 & \qswap     & \gate{A} & \qw \\
	\lstick{P}	      & \ghost{f}             & \qswap\qwx & \rstick{F} \qw \gategroup{2}{2}{3}{3}{.9em}{--} &  \\
}
\ee
where
\be
A = \bigotimes_{k=0}^{n-1}A^k,
\ee
and the $A^k$ are the parties' CPTP maps. This evolution was calculated in Appendix \ref{sec:proof_validity} to be the CPTP map
\be
G_A = \sum_{i,x,x'} U_{f(x)}A_i^T\ket{x}\bra{x'} {A_i^\dagger}^T U_{f(x')}\otimes \ket{x}\bra{x'}.
\ee
To simulate it, we need a purification of the CPTP maps $A^k$, that is, unitaries $U_{A^k}$ such that the Kraus operators $A^k_{i_k}$ are recovered from them via the formula
\be
A^k_{i_k} = \de{\bra{i_k}\otimes \id}U_{A^k}\de{\ket{0}\otimes \id}.
\ee
These unitaries are not unique, but any of them will work. Let then
\be
R = \bigotimes_{k=0}^{n-1} U_{A^k},
\ee
reordered such that all the ancillas are first. Then the circuit
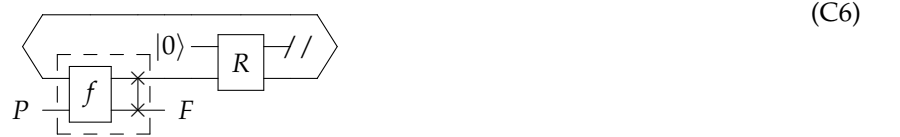
\begin{equation}
\Qcircuit @C=1em @R=0.3em @!R {
\preselection{2}{1.2} & \qw & \qw & \qw & \qw & \qw & \qw & \qw \postselection{2}{1.2} \\
		      &  &  &  & \lstick{\ket{0}\!\!} & \multigate{1}{R} & \qw {\ \ }{/} {/} & \\
&  \multigate{1}{f} &\qswap & \qw & \qw & \ghost{R} &\qw & \qw  \\
\lstick{P} & \ghost{f} & \qswap\qwx & \rstick{F} \qw \gategroup{3}{2}{4}{3}{.9em}{--}
}
\end{equation}
induces the same evolution from $P$ to $F$ as \eqref{eq:circuit_detn_cptp}, and so does the following analogue of circuit \eqref{eq:wdetcausalcircuitgeneral}:
\be
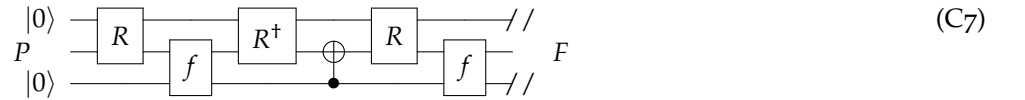
\label{eq:circuit_detn_cptp_ordered}
\Qcircuit @C=1em @R=0.3em @!R {
\lstick{\ket{0}} & \multigate{1}{R} & \qw              & \multigate{1}{R^\dagger} & \qw       & \multigate{1}{R} & \qw       & \qw{\ \ }{/}{/} & \\
\lstick{P\quad} & \ghost{R}        & \multigate{1}{f} & \ghost{R^\dagger}        & \targ     & \ghost{R} & \multigate{1}{f} & \rstick{\quad F} \qw \\
 \lstick{\ket{0}} & \qw & \ghost{f}        & \qw                      & \ctrl{-1} & \qw & \ghost{f} & \qw{\ \ }{/}{/} & \\
}
\ee
To see that it works, let us build the direct CJ representation of circuit \eqref{eq:circuit_detn_cptp_ordered}, ignoring for now the partial traces:
\be\label{eq:circuit_cj_general}
\Ket{S} = \sum_{xyz} \ket{x} (\id \otimes \ketbra{z}) R (\id \otimes U_{f(y)}) R^\dagger (\id \otimes \ketbra{y}) R \ket{0}\ket{x}U_{f(z)}U_{f(y)}\ket{0},
\ee
The crucial thing to notice is that the generalisation of property \eqref{eq:mysterious_property} holds: $f(y) \neq f(z)$ implies that
\be\label{eq:mysterious_property_n}
(\id \otimes \ketbra{z}) R (\id \otimes U_{f(y)}) R^\dagger (\id \otimes \ketbra{y}) = 0
\ee
and
\be\label{eq:mysterious_property_n_dagger}
(\id \otimes \ketbra{z}) R (\id \otimes U_{f(z)}) R^\dagger (\id \otimes \ketbra{y}) = 0
\ee
To see that this is true, first notice that equation \eqref{eq:mysterious_property_n_dagger} is just a relabelled adjoint of equation \eqref{eq:mysterious_property_n}, so it is enough to prove that $f(y) \neq f(z)$ implies equation \eqref{eq:mysterious_property_n}. We shall proceed by cases. First consider the case where $f(y) = 0^n$. Then $U_{f(y)} = \id$, and the \emph{lhs} of the above expression reduces to $\id \delta_{zy}$. If $f(y) \neq f(z)$, in particular we have $y \neq z$, which implies that $\delta_{zy} = 0$. 

Consider now the case where the first bit of $f(y)$ is one. From Appendix \ref{sec:proof_validity}, we know that this implies that $f(y) = 10^{n-1}$, and that $y = 00^{n-2}1$ or $y = 10^{n-2}1$. Substituting these values in equation \eqref{eq:mysterious_property_n}, we get
\be
(\id \otimes \ketbra{z_0}) U_{A^0} (\id \otimes X) U_{A^0}^\dagger (\id \otimes \ketbra{y_0})\delta_{1z_{n-1}}\prod_{i=1}^{n-2}\delta_{0z_i} = 0,
\ee
which is satisfied for all $z$ different than $00^{n-2}1$ or $10^{n-2}1$. Since these are precisely the values of $z$ for which $f(y) \neq f(z)$, the property also holds for this case. Since furthermore the function $f$ is translation invariant, the translated argument applies for all cases where $f(y) \neq 0$, and the claim follows. 

Substituting then $f(y)$ with $f(z)$ in equation \eqref{eq:circuit_cj_general} we get
\begin{align}
\Ket{S} &= \sum_{xz} \ket{x} (\id \otimes \ketbra{z}) R (\id \otimes U_{f(z)}) \ket{0}\ket{x}\ket{0} \\
	&= \sum_{z} (\id \otimes U_{f(z)})R^{T_P} \ket{0}\ket{z}\ket{z}\ket{0},
\end{align}
where $\cdot^{T_P}$ is a partial transposition over $P$. Applying the partial trace over both ancillas gives us then
\begin{align}
\tr_\text{anc}\KetBra{S} &= \sum_{izz'} (\bra{i} \otimes U_{f(z)})R^{T_P} \ketbra{0}\otimes \ket{z}\bra{z'}{R^{T_P}}^\dagger (\ket{i} \otimes U_{f(z')}) \otimes \ket{z}\bra{z'} \\
  &= \sum_{izz'} U_{f(z)} \Big((\bra{i} \otimes \id)R^{T_P}(\ket{0}\otimes \id) \Big) \ket{z}\bra{z'} \Big((\bra{0} \otimes \id){R^{T_P}}^\dagger(\ket{i}\otimes \id)\Big) U_{f(z')} \otimes \ket{z}\bra{z'} \\
  &= \sum_{izz'}  U_{f(z)} A_i^T \ket{z}\bra{z'} {A_i^T}^\dagger U_{f(z')} \otimes \ket{z}\bra{z'}
\end{align}
\qed

\end{document}